# Theory of Gelation
## Post−Gelation Behavior

*Kazumi Suematsu*
Institute of Mathematical Science
Ohkadai 2-31-9, Yokkaichi, Mie 512-1216, JAPAN
Fax: +81 (0) 593 26 8052, E-mail address: suematsu@m3.cty-net.ne.jp

## Summary

Within the framework of the random distribution assumption of cyclic bonds, the preceding theory of gelation is extended to mixing systems with various functionalities. To examine the validity of the assumption, the theory is applied to experimental data in polyurethane network formation, the result showing the soundness of the theory for the prediction of gel points and gel fraction.

## 1. Introduction

There remain some unsolved problems in polymer science. Of those, the determination of the gel point of branched polymer solutions has been one of the most attracting subjects for more than 60 years. It has been shown that the gel point varies widely from systems to systems, often deviating markedly from the ideal values [1]. It was pointed out earlier that the discrepancy between the theoretical point and observed ones can be ascribed to the neglect of cyclization [2]. Based on the findings of the early researchers, the author has pursued so far the general theory of gelation that includes cyclization effects [3].

In this report, the author will extend the preceding theory to mixing systems. The essence of our approach is based on the following four premises (three principles and one assumption):

### Three Principles

(i) *The gel point is divided into the two terms*:
$$D_c = D(inter) + D(ring). \qquad (1)$$

(ii) *The total ring concentration, $[\Gamma]$, is independent of the initial monomer concentration, $C$; it is a function of D (the extent of reaction) alone.*

(iii) *Branched molecules behave ideally at $C = \infty$.*

### One Assumption

(iv) Assumption I: *Cyclic bonds distribute randomly over all bonds.*

The premise (i) is a mathematical theorem. The premise (ii) is based on the following formal solution of $[\Gamma]$:

$$[\Gamma] \propto C \sum_{j=1}^{\infty} \int_D \frac{(v_{R_j}/v_L)}{1+(v_R/v_L)} dD, \qquad (2)$$

which is common to all model systems, where $v_{R_j}$ denotes the velocity of $j$-ring formation and $v_R = \sum_j^{\infty} v_{R_j}$, and $v_L$ that of intermolecular reaction. Experiments have shown that the relative velocity, $v_{R_j}/v_L$, is a decreasing function of the initial monomer concentration, $C$, which implies that eq. (2) is of the form: $[\Gamma] \propto C \int 1/(1+t) dD$, a function rapidly approaching a constant with increasing $t$ ($t$ is a representation of the concentration of reactants). This tells us that $[\Gamma]$ is constant at high concentration, which, in fact, has been verified rigorously for $f = 2$ [3]. The premise (iii) can not be verified immediately by experimental methods, but has solid physical foundation: (a) experiments have shown that $v_{R_j}/v_L \to 0$ as $C \to large$, so that the production of rings becomes negligible at high concentration; (b) all excluded volume effects vanish rigorously at $C = \infty$ [3], [4], since, as $C \to \infty$, the monomer density becomes infinite and the notion of atomic radii vanishes.

In contrast to the above three principles, the premise (iv) (Assumption I) is an approximation. Assumption I, however, greatly simplifies the problem. It reduces otherwise an inherently insoluble problem of gelation to an elementary mathematical exercise, giving the equality $D(inter) = D_{co}$.

Eq. (1) then becomes

$$D_c = D_{co} + p_R, \qquad (3)$$

with $D_{co}$ being the Flory point, and $p_R$ the fraction of cyclic bonds to all possible bonds and can be equated with $D(ring)$ [3].

In this report, the author generalizes, within the framework of the above three principles and one assumption, the preceding theory to include mixing systems of multifunctionalities.

## 2. Theoretical
### 2-1) R-A$_f$ Model
### Gel Point

Suppose that a reaction system is comprised of a mixture, $\{f_i M_i\}$, of branching units, with $f_i$ being the functionality and $M_i$ the mole number. Let $\chi_i = f_i M_i / \sum_i f_i M_i$ be the fraction of functional units (FU's) belonging to the $i$th branching unit. Let $D$ be the extent of reaction of all FU's, $J$ be the number of FU's to form a junction point. Let $X$ and $Z$ be sets of all bonds and cyclic bonds, respectively. Each ring possesses only one excess (cyclic) bond. Under *Assumption I*, the probability, $\alpha$, that a branching unit leads to the next branching unit to form a network is equal to the product of the fraction of reacted functional units and the fraction of intermolecular bonds. Hence

$$\alpha = \sum_i \chi_i \sum_{k=0}^{f_i-1} k \binom{f_i-1}{k} D^k (1-D)^{f_i-1-k} \sum_{\ell=0}^{J-1} \ell \binom{J-1}{\ell} (1-P(Z|X))^{\ell} P(Z|X)^{J-1-\ell}. \qquad (4)$$

where $P(Z|X) = P(Z \cap X)/P(X)$ is the conditional probability that a randomly chosen bond is a cyclic bond. Thus the quantity, $1 - P(Z|X)$, represents the probability that a chosen bond is an intermolecular bond. The conditional probability term is essential, since, if a bond is a cyclic bond, it is impossible, according to the definition of a loop, for the bond to extend infinitely over subsequent branching units. Equally to the preceding case, one has

$$P(Z|X) = \frac{p_R}{D}. \qquad (5)$$

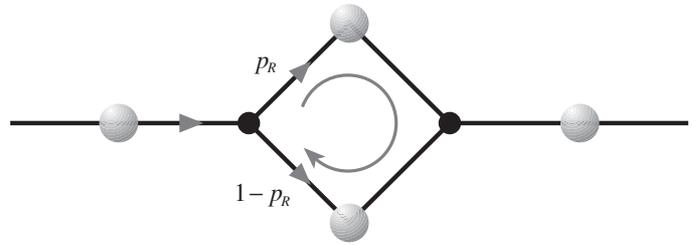

Fig. 1. Representation of a growing branch with one loop (◓: branching unit; ●: junction point).

The gelation occurs at $\alpha = 1$, and eq. (4) gives

$$D_c = 1/(J-1)(\langle f_w \rangle - 1) + p_R, \qquad (6)$$

with $\langle f_w \rangle = \sum_i f_i \chi_i$. The first term of eq. (6) represents the ideal gel point and the second term the correction term due to cyclization. Our task is then to find the mathematical expression of $p_R$ for this mixing system.

As every ring possesses only one excess bond and $(J-1)$ branches arise by the merger of $J$

FU's, $p_R$ should have the form:

$$p_R = \frac{J[\Gamma]}{(J-1)C_f}, \qquad (7)$$

$C_f$ denoting the FU concentration defined by $C_f = \sum_i f_i M_i / V$ ($V$ is a system volume), while $[\Gamma]$ having the formal solution:

$$[\Gamma] = \frac{fC}{J} \sum_{j=1}^{\infty} \int_0^D \frac{(v_{R_j}/v_L)}{1+(v_R/v_L)} dD. \qquad (8)$$

Experiments have shown that $(v_{R_j}/v_L)$ is in inverse proportion to the concentration of reactants, so that as $C \to \infty$, the relative frequency of cyclization becomes negligible. Hence, at high concentration, eq. (8) can be approximated in the form:

$$[\Gamma] \doteq \frac{fC}{J} \sum_{j=1}^{\infty} \int_0^D (v_{R_j}/v_L) dD. \qquad (9)$$

This means that at sufficiently high concentration, $[\Gamma]$ can be equated with the limiting solution of $C \to \infty$:

$$[\Gamma] \doteq [\Gamma]_{C \to \infty}. \qquad (10)$$

At $C = \infty$, eq. (8) has an analytic solution of the form [5]:

$$[\Gamma]_{C \to \infty} = \sum_{j=1}^{\infty} \varphi_j \left[ (J-1)(\langle f_w \rangle - 1) D \right]^j / 2j, \qquad (11)$$

$\langle f_w \rangle$ being the mean weight functionality as defined above.

Because of the relationship (10), we can expect that eq. (11) is a good estimate of the ring concentration in actual branching processes. Unfortunately, eq. (11) breaks down as soon as the ideal gel point, $D_{co} = 1/(J-1)(\langle f_w \rangle - 1)$, is passed, so eq. (11) cannot directly be linked with eqs. (6) and (7). To resolve this problem, we make use of a linear approximation of eq. (8), or equivalently of eq. (11) [3]. Recall that the ring concentration has the formal solution at $D_c$:

$$[\Gamma] = \mathcal{C}(D_c) = \frac{fC}{J} \sum_{j=1}^{\infty} \int_0^{D_c} \frac{(v_{R_j}/v_L)}{1+(v_R/v_L)} dD, \qquad (8')$$

while by eq. (10), at high concentration $[\Gamma]$ is a function of $D$ alone, so one can expand eq. (8') with respect to $D_{co}$

$$\mathcal{C}(D_c) = \mathcal{C}(D_{co}) + \frac{\mathcal{C}'(D_{co})}{1!}(D_c - D_{co}) + \cdots,$$

giving

$$[\Gamma] \cong \sum_{j}^{\infty} \varphi_j / 2j + \frac{(J-1)(\langle f_w \rangle - 1)}{2} \sum_{j}^{\infty} \varphi_j (D_c - D_{co}), \tag{12}$$

which, now, can be applied to the new territory, $D_{co} \leq D \leq D_c$. Eq. (12) has sound mathematical basis because of the relationship of eq. (10). From eqs. (6), (7) and (12), one has

$$D_c = \frac{1}{(J-1)(\langle f_w \rangle - 1)} \left\{ \frac{1 - \frac{J(\langle f_w \rangle - 1)}{2} \sum_{j}^{\infty} (1 - 1/j) \varphi_j \gamma_f}{1 - \frac{J(\langle f_w \rangle - 1)}{2} \sum_{j}^{\infty} \varphi_j \gamma_f} \right\}, \tag{13}$$

where $\gamma_f = 1/C_f$ is the reciprocal of the initial FU concentration. For a single component system, $\langle f_w \rangle \to f$ and $\gamma_f \to \gamma/f$, and one recovers the preceding result. Thus eq. (13) is an extension of the previous expression.

**Post-gelation**

We confine ourselves to a mixture of monomers with $f \geq 2$. Let $Q$ be the probability that a chosen branch emanating from a monomer is finite [6]-[8]. Following the aforementioned Assumption I, $Q$ satisfies the recurrence relation:

$$Q = 1 - D + D \left\{ \sum_{i} \chi_i \left( \frac{p_R}{D} + \left(1 - \frac{p_R}{D}\right) Q^{f_i - 1} \right) \right\}^{J-1}. \tag{14}$$

It is easy to show that eq. (14) has two solutions, $Q = 1$ and

$$(D - p_R) \sum_{i} \chi_i \left( Q^{f_i - 2} + Q^{f_i - 3} + \cdots + 1 \right) \left( X^{J-2} + X^{J-3} + \cdots + 1 \right) = 1, \tag{15}$$

with

$$X = \sum_{i} \chi_i \left( \frac{p_R}{D} + \left(1 - \frac{p_R}{D}\right) Q^{f_i - 1} \right).$$

The assignment of branching units can be made by the binomial expansion:

$$\sum_{i} \omega_i (1 - Q + Q)^{f_i} = \sum_{i} \omega_i \left\{ \binom{f_i}{0}(1-Q)^0 Q^{f_i} + \binom{f_i}{1}(1-Q)^1 Q^{f_i - 1} + \cdots \right\}, \tag{16}$$

with $\omega_i = \frac{m_i M_i}{\sum_i m_i M_i} = \frac{\chi_i m_i / f_i}{\sum_i \chi_i m_i / f_i}$ being the weight fraction of the $i$th monomer unit and $m_i$ the molecular mass. The first term of eq. (16) represents the weight fraction ($w_s$) of sol, the second term the fraction ($w_p$) of pendants, and the higher terms than the third the fraction of active network segments. Hence

$$w_s = 1 - w_g = \sum_{i} \omega_i Q^{f_i}, \tag{17}$$

$$w_p = \sum_{i} \omega_i \binom{f_i}{1}(1-Q)^1 Q^{f_i - 1}. \tag{18}$$

To calculate the post-gelation behavior, we extend the concept of cyclization in sol phase to gel phase; namely, we apply the foregoing linear approximation shown in eq. (12) to the gel phase and write

$$[\Gamma] \cong \sum_{j}^{\infty} \varphi_j / 2j + \frac{(J-1)(\langle f_w \rangle - 1)}{2} \sum_{j}^{\infty} \varphi_j (D - D_{co}), \qquad (19)$$

for $D_{co} \leq D \leq 1$. In this expression, it is assumed that all reactions are smooth and continuous beyond the gel point with no reaction anomaly. Substitution of eq. (19) into eq. (17), with the help of eqs. (7) and (15), gives a solution of $w_g$ as functions of $D$, $\kappa$ and $\gamma_f$.

**$Q = 1$, Critical Case**

Set $Q = 1$ and eq. (15) gives the critical condition, $D_c = 1/(J-1)(\langle f_w \rangle - 1) + p_R$, in agreement with eq. (6).

**2-2) R-$A_g$ + R-$B_{f-g}$ Model**

**Gel Point**

Let there be a mixing system comprising of two different types of monomer units, $\{g_i M_{A_i}\}$ and $\{(f-g)_j M_{B_j}\}$, where $M_{A_i}$ and $M_{B_j}$ are the mole numbers of the A type and the B type monomers, respectively, and $g_i$ and $(f-g)_j$ are the corresponding functionalities. Let $J$ be the total number of FU's to form a junction point on which the two types of the FU's are arranged alternately. By the nature of the R-$A_g$ + R-$B_{f-g}$ model, $J$ must be an even integer and $J-1$ branches arise as a result of the merger of $J/2$ A type FU's and $J/2$ B type FU's. Thus, the probability, $\alpha$, of branching becomes

$$\alpha = D_A \left\{ \left(\tfrac{J}{2}-1\right)(1-P_{AA}(Z|X))(\langle g_w \rangle -1) + \tfrac{J}{2}(1-P_{AB}(Z|X))(\langle (f-g)_w \rangle -1) D_B \sum_{i=0}^{\infty} s^i \tfrac{J}{2}(1-P_{AB}(Z|X))(\langle g_w \rangle -1) \right\}$$

(20)

where $s = \left(\tfrac{J}{2}-1\right)(1-P_{BB}(Z|X))(\langle (f-g)_w \rangle -1) D_B$, the subscripts, AA, AB and BB denote cyclic formation via respective bond species, and the subscript, $w$, inside the bracket $\langle \cdots \rangle$ the weight average quantity. For $\alpha = 1$, eq. (20) gives the gelation condition:

$$D_A \left\{ \left(\tfrac{J}{2}-1\right)(1-P_{AA}(Z|X))(\langle g_w \rangle -1) + \frac{\left(\tfrac{J}{2}\right)^2 (1-P_{AB}(Z|X))^2 (\langle g_w \rangle -1)(\langle (f-g)_w \rangle -1) D_B}{1-\left(\tfrac{J}{2}-1\right)(1-P_{BB}(Z|X))(\langle (f-g)_w \rangle -1) D_B} \right\} = 1.$$

(21)

Unfortunately, the solution is much intricate and difficult to use. So, here we examine the $J = 2$ case only, for which eq. (21) reduces to

$$(\langle g_w \rangle -1)(\langle (f-g)_w \rangle -1)(1-P_{AB}(Z|X))^2 D_A D_B = 1. \qquad (21')$$

Substituting $P_{AB}(Z|X) = p_R / D_A$ into eq. (21'), we have the critical condition:

$$D_c = \sqrt{\kappa/(\langle g_w\rangle-1)(\langle(f-g)_w\rangle-1)} + p_R, \tag{22}$$

with $D_c$ representing the extent of reaction of A FU's. Eq. (22) is the $R-A_g+R-B_{f-g}$ model version of eq. (6). Since $p_R$ is the ratio of cyclic bonds to all possible bonds, it may be written in the form:

$$p_R = \frac{[\Gamma]}{C_{f,A}} = (1+\kappa)[\Gamma]\gamma_f, \tag{23}$$

with $C_{f,A} = \sum_i g_i M_{A_i}/V$ being the concentration of A FU's, $\kappa = \sum_j (f-g)_j M_{B_j}/\sum_i g_i M_{A_i}$, the molar ratio of B FU's to A FU's (we define $\kappa \geq 1$), and

$$\gamma_f = V/\left(\sum_i g_i M_{A_i} + \sum_j (f-g)_j M_{B_j}\right), \tag{24}$$

the reciprocal of the total FU concentration.

$p_R$ as a function of $D_c$ can be obtained in the same manner as in eq. (12). We write

$$[\Gamma] = C_{f,A} \sum_{j=1}^{\infty} \int_0^{D_c} \frac{(v_{R_j}/v_L)}{1+(v_R/v_L)} dD, \tag{25}$$

for which eq. (25) has the asymptotic solution of $C\to\infty$:

$$[\Gamma]_{C\to\infty} = \sum_{j=1}^{\infty} \varphi_j \left[(\langle g_w\rangle-1)(\langle(f-g)_w\rangle-1)D^2/\kappa\right]^j / 2j. \tag{26}$$

With the relationship, $[\Gamma] \cong [\Gamma]_{C\to\infty}$, at high concentration in mind, expand eq. (25) with respect to $D = D_{co}$ and substitute into eq. (23) to get

$$p_R = (1+\kappa)\left\{\sum_{j=1}^{\infty} \varphi_j/2j + \frac{1}{D_{co}}\sum_{j=1}^{\infty} \varphi_j(D_c - D_{co})\right\}\gamma_f. \tag{27}$$

Substituting eq. (27) into eq. (22), we have the gel point expression for the mixing system of the $R-A_g+R-B_{f-g}$ model $(J=2)$:

$$D_c = D_{co}\left\{\frac{1-\mathcal{K}\sum_j(1-1/2j)\varphi_j\gamma_f}{1-\mathcal{K}\sum_j\varphi_j\gamma_f}\right\}. \tag{28}$$

Here $\mathcal{K} = (1+\kappa)/D_{co}$ and $D_{co} = \sqrt{\kappa/(\langle g_w\rangle-1)(\langle(f-g)_w\rangle-1)}$. Eq. (28) just corresponds to the following transformation of the homogeneous system [3]:

$$\gamma \rightarrow \gamma_f;$$

*functionality* $\rightarrow$ *weight average functionality*.

### Post−gelation ($J = 2$)

We consider the mixture of $f \geq 2$ monomers. Let $Q_A$ be the probability that a branch emanating from an A type monomer unit is finite, and $Q_B$ the corresponding probability for a B type unit.

Then $Q_A$ and $Q_B$ satisfy

$$Q_A = 1 - D_A + D_A \sum_j \chi_{Bj} \left( \frac{p_R}{D_A} + \left(1 - \frac{p_R}{D_A}\right) Q_B^{(f-g)_j - 1} \right);$$

$$Q_B = 1 - D_B + D_B \sum_i \chi_{Ai} \left( \frac{p_R}{D_A} + \left(1 - \frac{p_R}{D_A}\right) Q_A^{g_i - 1} \right), \quad (29)$$

with $\chi_{Ai} = g_i M_{A_i} / \sum_i g_i M_{A_i}$ and $\chi_{Bj} = (f-g)_j M_{B_j} / \sum_j (f-g)_j M_{B_j}$. The binomial expansion of the probabilities, $\{Q\}$, is

$$\sum_i \omega_{A_i}(1 - Q_A + Q_A)^{g_i} + \sum_j \omega_{B_j}(1 - Q_B + Q_B)^{(f-g)_j} = \sum_i \omega_{A_i} \left\{ \binom{g_i}{0} Q_A^{g_i} + \binom{g_i}{1}(1 - Q_A) Q_A^{g_i - 1} + \cdots \right\} +$$

$$\sum_j \omega_{B_j} \left\{ \binom{(f-g)_j}{0} Q_B^{(f-g)_j} + \binom{(f-g)_j}{1}(1 - Q_B) Q_B^{(f-g)_j - 1} + \cdots \right\}. \quad (30)$$

The respective first terms represent the weight fraction of sol:

$$w_s = 1 - w_g = \sum_i \omega_{A_i} Q_A^{g_i} + \sum_j \omega_{B_j} Q_B^{(f-g)_j}, \quad (31)$$

where $\omega_{A_i}$ and $\omega_{B_j}$ are the weight fractions of the $i$th A type and the $j$th B type monomer units, respectively.

**Special Solution I**

Let us consider a mixing system of $g_i = 2$ (for all $i$'s), $(f-g)_1 = 2$, $(f-g)_2 = 3$ and $J = 2$. This special system has an application to the polyurethane network formation of MDI (4,4'-diphenylmethane diisocyanate: $m_A = 250$) and LHT240 (polyoxypropylene triol: $\langle m_B \rangle = 708$). LHT240 is considered to be a mixture of diols and triols [9]:

There are conflicting views about chemical constituents of LHT240. Adam and coworkers [10] report that LHT240 is purely trifunctional, while Ilavsky and coworkers [9] report that it is a mixture of diols and triols. The difference of the views stems from the difference in the characterization of LHT240, especially in the estimation of OH content (see Table 1). Although which estimation is

Table 1. Chemical characterization of LHT240

|  | Adam et al [10] | Ilavsky et al [9] |
| --- | --- | --- |
| Number Average MW | 715 by SEC | 708 by VPO |
| OH Content (w/w %) | 7.14 | 6.94 |
| Calculated Functionality | $f - g = 3$ | $\langle (f-g)_n \rangle = 2.89$ |

SEC: Size exclusion chromatography
VPO: Vapour pressure osmometry

Table 2. Chemical constants of MDI-LHT240 mixture [3]

| | |
|---|---|
| Functionality | $g = 2$, $\langle (f-g)_n \rangle = 2.89$, $\langle (f-g)_w \rangle = 2.92$, $\chi_{B2} = 0.92$ |
| Mean Molecular Mass | $m_A = 250\,(MDI)$, $\langle m_B \rangle = 708\,(LHT240)$ |
| Flory Characteristic Constant | $C_F = 4.5$ |
| Standard Bond Length | $\ell = 1.36$ Å |
| Effective Bond Number | $\xi = 68$ |
| Relative Frequency | $\sum_{j=1}^{\infty} \varphi_j / 2j = 0.0272$, $\sum_{j=1}^{\infty} \varphi_j = 0.1056$ |

correct is indecisive at present, we use here the Ilavsky and coworkers estimation, since then the theory can reasonably explain all the observed points.

Let $\chi_{B2}$ be the mole fraction of the triols to the total alcohols as defined above. Eq. (29) yields

$$Q_A = 1 + \frac{\kappa\{\kappa - (1+\chi_{B2})(D-p_R)^2\}}{\chi_{B2}(D-p_R)^3}, \quad Q_B = \frac{\kappa - (D-p_R)^2}{\chi_{B2}(D-p_R)^2}. \tag{32}$$

Substituting into eq. (31), one has the weight fraction of gel:

$$w_g = 1 - \omega_A Q_A^2 - \omega_{B1} Q_B^2 - \omega_{B2} Q_B^3. \tag{33}$$

In Figs. (2) and (3), eq. (33) is plotted as a function of $\gamma_f$ and $\kappa$ at $D = 1$, respectively. The parameters employed here are those [3] of MDI-LHT240 (see Table 2). The relative cyclization frequency, $\varphi_j$, the only unknown parameter, is evaluated, based on the premise (iii), by the equation:

$$\varphi_j = \frac{d}{2\pi^{d/2}\ell^d N_A}\left[\Gamma\left(\tfrac{d}{2}\right) - \Gamma\left(\tfrac{d}{2}, \tfrac{d}{2v}\right)\right] \quad [mole/l], \tag{34}$$

where $d$ is dimension, $\ell$ is the standard bond length which can be equated with the C-N bond length (1.36 Å), $\Gamma\left(\tfrac{d}{2}, \tfrac{d}{2v}\right)$ is the incomplete Gamma function, and $v$ is the proportionality factor defined by $\langle r_j^2 \rangle_\Theta = v\ell^2 = C_F \xi j \ell^2$, with $C_F$ being the Flory characteristic constant in the $\Theta$ state and $\xi$ the effective bond number per a repeating unit of the polymer backbone [3].

The weight fractions, $\{\omega\}$, can be calculated using the molecular weights, $\{m\}$, of respective branching

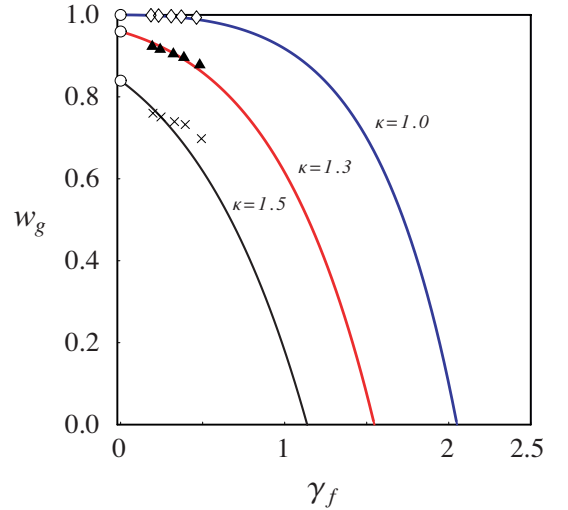

Fig. 2. $w_g$ vs $\gamma_f$ curves of the MDI-LHT240 mixture ($D = 1$).
Solid line: theoretical line by eq. (33);
◇ ($\kappa = 1.0$), ▲ ($\kappa = 1.3$), and × ($\kappa = 1.5$): experimental points by Ilavsky and Dusek;
○ : theoretical points by the ideal tree theory.
$\gamma_f$ is a function of $\kappa$ itself and was calculated by the formula:

$$\gamma_f = \frac{\langle (f-g)_n \rangle \langle m_{An} \rangle + \langle g_n \rangle \langle m_{Bn} \rangle \kappa}{1000 \rho v \langle g_n \rangle \langle (f-g)_n \rangle (1+\kappa)} \quad (l/mol),$$

where the subscript $n$ denotes the number average, $\rho$ is the density of the polymeric materials and $v$ the volume fraction.

units. For the diols ($m_{B1}$) and the triols ($m_{B2}$) in question, the observed heterogeneity index, $\langle m_w \rangle / \langle m_n \rangle = 1.03$ [10] provides two possible solutions, $\{m_{B1}, m_{B2}\} = \{\{359, 751\}, \{1057, 665\}\}$.

Not surprisingly, graphical examination showed that these two solutions gave almost identical $w_g - \kappa$ curves when applied to eq. (33). For this reason, we mention only the result for the solution, $\{m_A, m_{B1}, m_{B2}\} = \{250, 1057, 665\}$.

The solid lines in Fig. 2 show the theoretical ones; the symbols ($\diamondsuit$: $\kappa = 1.0$; $\blacktriangle$: $\kappa = 1.3$; $\times$: $\kappa = 1.5$) are the experimental points by Ilavsky and Dusek [9]; the open circles ($\bigcirc$) represent the theoretical values predicted by the ideal tree theory ($\gamma_f = 0$) with no rings and no excluded volume effects. By the aforementioned reasons (Section I), we expect that all the data should converge on these limiting cases ($\bigcirc$), as $\gamma_f \to 0$. Both the theory and the experiments show that $w_g$ is less than unity at $D = 1$, suggesting that the permanent sol discussed precedently [3] is produced more abundantly with increasing $\gamma_f$.

Fig. 3 shows the $\kappa$ dependence of $w_g$. The solid line is the theoretical one calculated by eq. (33), and the broken line is the ideal case of $p_R = 0$; open circles ($\bigcirc$) are observed points. Agreement between the theory and the experiments is satisfactory.

The gelation process by the reaction of HMD (hexamethylene diisocyanate: $m_A = 168$) and LHT240 was investigated by Durand and coworkers [10] under the condition without solvent. They observed gel fractions as a function of $\kappa$ (molar ratio of B FU's to A FU's). The chemical parameters of this system are shown in Table 3. Making use of these values, the theoretical line is drawn in Fig. 4 (solid line). Experimental points ($\times$) are Durand and coworkers'. Agreement between the experiment and the theory is satisfactory.

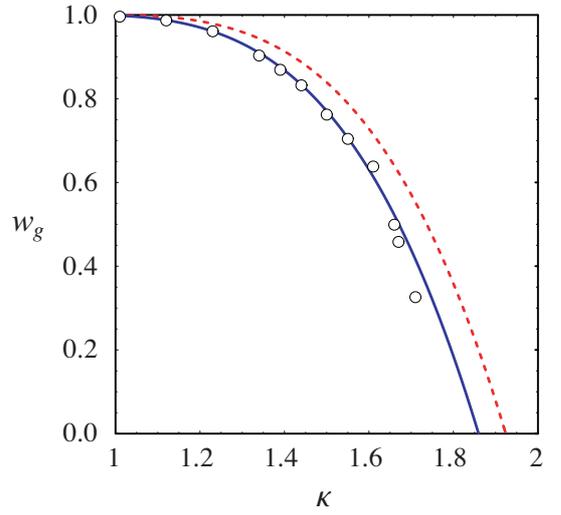

Fig. 3. $w_g$ vs $\kappa$ curves of the MDI-LHT240 mixture ($D = 1$).
Dotted line: theoretical line by the ideal tree theory; solid line: theoretical line by eq. (33); $\bigcirc$: experimental points by Ilavsky and Dusek.

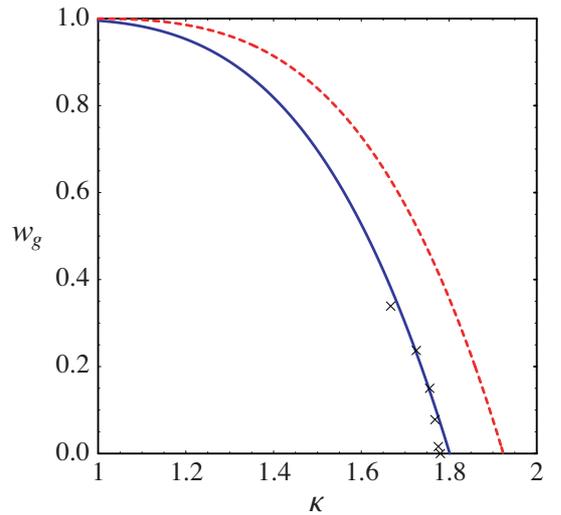

Fig. 4. $w_g$ vs $\kappa$ curve of the HMD-LHT240 mixture ($D = 1$).
Dotted line: theoretical line by the ideal tree theory; solid line: theoretical line by eq. (33); $\times$: experimental points by Durand and coworkers.

### Special Solution II

Consider the homogeneous case of $g_i = 3$ (for all $i$'s), $(f - g)_j = 2$ (for all $j$'s) and $J = 2$. This system is a particular case of $\chi_{B2} = 1$ of Special Solution I. Using the boundary condition, $0 \le D_c \le 1$,

Table 3  Chemical constants of HMD-LHT240 mixture

| | |
|---|---|
| Functionality | $g = 2$, $\langle (f-g)_n \rangle = 2.89$, $\langle (f-g)_w \rangle = 2.92$, $\chi_{B2} = 0.92$ |
| Mean Molecular Mass | $m_A = 168\,(HMD)$, $\langle m_B \rangle = 708\,(LHT240)$ |
| Flory Characteristic Constant | $C_F = 4.5$ |
| Standard Bond Length | $\ell = 1.36\,\text{Å}$ |
| Effective Bond Number | $\xi = 41$ |
| Relative Frequency | $\sum_{j=1}^{\infty}\varphi_j/2j = 0.0583$, $\sum_{j=1}^{\infty}\varphi_j = 0.2267$ |

eq. (28) gives the critical-dilution condition:

$$\gamma_f \leq \gamma_{f,c} = \frac{1-D_{co}}{(1+\kappa)\sum_j^{\infty}(-1+1/D_{co}+1/2j)\varphi_j}, \qquad (35)$$

with $D_{co} = \sqrt{\kappa/2}$. In order for the gelation to occur, $\gamma_f$ must be less than the above critical value, $\gamma_{f,c}$. The weight fraction of gel is given by

$$w_g = 1 - \omega_A \left\{1 + \frac{\kappa\left(\kappa - 2(D-p_R)^2\right)}{(D-p_R)^3}\right\}^2 - \omega_B \left\{-1 + \frac{\kappa}{(D-p_R)^2}\right\}^3. \qquad (36)$$

Eq. (35) is applied to Budinski-Simendic and coworkers' experiments [11] for the polyurethane network made from TI (tris-4-isocyanatophenylthiophosphate: $m_A = 465$) and PD (polyoxypropylenediol: $m_B = 970$). With the help of the parameters in Table 4, eq. (35) can be plotted as a function of $\kappa$ (solid line; Fig. 5 & Fig. 6). The symbols (○) denote the experimental points by Budinski-Simendic and coworkers. Agreement between the theory and the observations is excellent. It is important to notice that the classical tree theory (broken line) cannot explain these observed values.

Table 4. Chemical constants of TI-PD mixture [11]

| | |
|---|---|
| Functionality | $g = 3$, $f - g = 2$ |
| Mean Molecular Mass | $m_A = 465\,(TI)$, $m_B = 970\,(PD)$ |
| Flory Characteristic Constant | $C_F = 4.3$ |
| Standard Bond Length | $\ell = 1.36\,\text{Å}$ |
| Effective Bond Number | $\xi = 98$ |
| Relative Frequency | $\sum_{j=1}^{\infty}\varphi_j/2j = 0.0169$, $\sum_{j=1}^{\infty}\varphi_j = 0.0680$ |

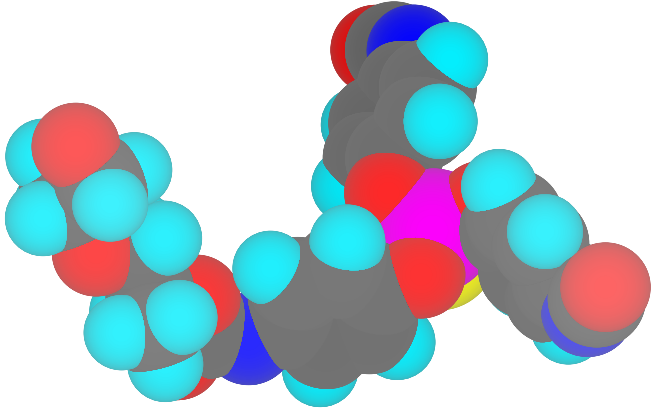

Fig. 5. Graphical representation of the repeating unit of the TI-PD polymer

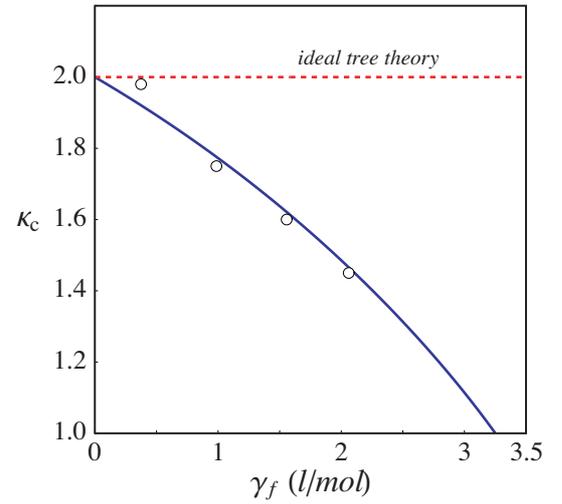

Fig. 6. $\gamma_f$ dependence of $\kappa_c$ for the TI-PD mixture. Dotted line: prediction of the ideal tree theory; solid line: prediction of eq. (35); ○: experimental points by Budinski-Simendic and co-workers.

In Fig. 7, eq. (36) is plotted as a function of $D$: the solid line (—) corresponds to $v = 1$ and the broken line (– –) $v = 0.2$ ($v$ denotes the volume fraction of polymer). The dotted line (···) shows the ideal tree theory with no rings. To date, within our knowledge, there are no experimental observations corresponding to the dilution regime of $v = 0.2$.

The results of Figs. (3)-(6) support the soundness of the theory. We must bear in mind, however, that Assumption I is an approximation. The experimental determination of gel points is a difficult task, so theoretical errors may be hidden behind experimental errors. For this reason, it will be necessary to scrutinize further the validity of Assumption I by extensive experimental observations.

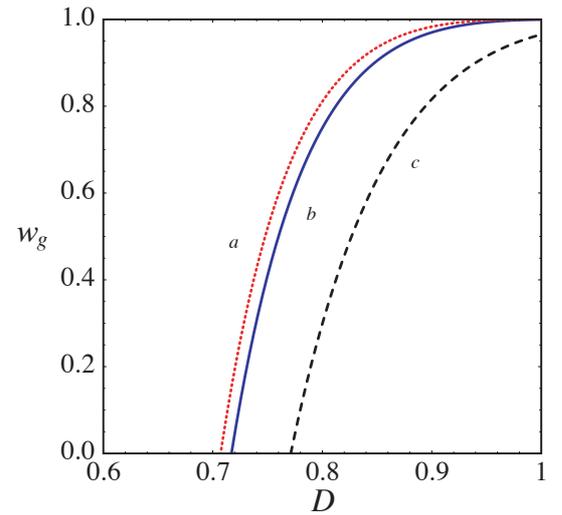

Fig. 7. $D$ dependence of $w_g$ for the TI-PD mixture for $\kappa = 1$.
(a) Dotted line: prediction of the ideal tree theory $(\gamma_f = 0)$; (b) solid line: prediction of eq. (36) for $v = 1$ $(\gamma_f = 0.278)$; (c) broken line: prediction of eq. (36) for $v = 0.2$ $(\gamma_f = 1.39)$.

## 4. Conclusion

The theory of gelation was applied to the observed data in polyurethane networks. Good agreement was found between the theory and the observations, both in the predictions of gel points and gel fractions.